\documentclass[a4paper,11pt]{article}
\pdfoutput=1 

\usepackage{jcappub} 

\usepackage{amsmath}
\usepackage{amsfonts,amssymb}
\usepackage{hyperref}
\usepackage{graphicx}
\usepackage[utf8]{inputenc}
\usepackage{xcolor}
\usepackage{bm}

\hypersetup{
    colorlinks, 
    linktoc=all, 
    linkcolor=blue,  
  citecolor=hyptxt,
  urlcolor=blue}
  \hypersetup{
  citebordercolor=,
  filebordercolor=green,
  linkbordercolor=blue
}
\definecolor{hyptxt}{rgb}{0.7, 0.4, 0.9}

\title{\boldmath Testing the Equivalence Principle with Strong Lensing Time Delay Variations}

\author[a,b,c]{Leonardo Giani}
\author[a,b]{and Emmanuel Frion}

\affiliation[a]{Department of Physics, Universidade Federal do Esp\'irito Santo,\\ Avenida Fernando Ferrari 514, 29075-910 Vit\'oria, Esp\'irito Santo, Brazil}
\affiliation[b]{N\'ucleo Cosmo-ufes and PPGCosmo, Universidade Federal do Esp\'irito Santo,\\ Avenida Fernando Ferrari 514, 29075-910 Vit\'oria, Esp\'irito Santo, Brazil}
\affiliation[c]{Institut f\"{u}r Theoretische Physik, Ruprecht-Karls-Universit\"{a}t Heidelberg, Philosophenweg 16, D-69120 Heidelberg, Germany}

\emailAdd{giani@thphys.uni-heidelberg.de}
\emailAdd{emmanuel.frion@cosmo-ufes.org}

\abstract{
 Strong lensing time delay measurements provide a valuable and almost model-independent tool for cosmological investigations. In this work we recognize that they also carry information on the strength of the gravitational coupling at the redshift of the lens, and thus could be in principle used to test the equivalence principle on extragalactic scales. For the case of an extended lens with a static mass distribution we explicitly derive an analytical relation between $\dot{G}/G$ and the relative variation of the time delay. For illustrative purpose, we apply our formula to the light curves of multiple images of the quasar DES J0408-5354 and simulated ones,  which results in weak constraints on the variation of $\dot{G}/G$ of order $10^{-1}-10^{-2} \, yr^{-1}$ in the best scenario. Finally, we briefly discuss how those constraints can be improved in the next future.
}

\begin{document}
\maketitle
\flushbottom

\section{Introduction}
Cosmology is a very vivacious research field where several branches of physics combine with the ambitious goal of describing the evolution of our Universe. In the last decades, thanks to the joint effort of several researchers, the state-of-the-art observations brought us into the era of \textit{precision} Cosmology \cite{Primack:2004gb}.  Along with this increasing precision appeared tensions on the  measured value of cosmological parameters, see for example Refs.~\cite{Verde:2019ivm,Freedman:2017yms,Douspis:2018xlj,Kazantzidis:2018rnb,Bernal:2016gxb}, signaling maybe the necessity of new physics beyond the Standard Model \cite{Riess:2019cxk}. To deal properly with this issue, it is of utmost importance to develop observational techniques which are \textit{model-independent}, i.e. that minimize the number of assumptions required to infer the cosmological parameters from the data, see for example Ref.~\cite{Amendola:2019laa}.  

An observable whose measure is model-independent is the so-called \textit{redshift drift}, defined as the time variation of the redshift of a source due to the Hubble flow, which thus relies only on the assumptions of homogeneity and isotropy.
It was proposed for the first time by Sandage and McVittie in 1962 \cite{SandageMcVittie}. At that time, they concluded that it was impossible to detect such an effect in a reasonable time span, but is nowadays a promising observational task for collaborations like SKA or ELT. For this reason, probing cosmology via redshift drift is an active research field, see for example Refs.~\cite{Alves:2019hrg,Klockner:2015rqa,Melia:2016bnb,Bolejko:2019vni,Martinelli_2012,Quartin:2009xr}. 

Another interesting and \textit{almost} model-independent observable is the time delay of incoming signals from multiple images in a strongly gravitational lensed system, which is due to the different paths that the photons emitted by the source have to travel because of the presence of the lens, see for example \cite{Weinberg:2008zzc}. Within the COSMOGRAIL \footnote{\href{http://www.cosmograil.org}{http://www.cosmograil.org}} program, the H0LiCOW collaboration \footnote{\href{https://shsuyu.github.io/H0LiCOW/site/index.html}{https://shsuyu.github.io/H0LiCOW/site/index.html}} employed time delays measured over the last decade to constrain the value of the cosmological parameter $H_0$ to a few percents level in a model-independent way \cite{Suyu:2016qxx,Wong:2019kwg,Bonvin:2016crt}, with competitive precision with respect to other cosmological probes. Strong lensing time delay measurements can also be used to put constraints on the Post-Newtonian parameter $\gamma_{PPN}$, as discussed in Refs.~\cite{Collett:2018gpf,Yu:2018slt,Jyoti:2019pez,Yang:2020eoh}.  Furthermore, with optimistic assumptions on the surveys, in the next years the precision of observations will be enough to provide a smoking gun for Dark Energy \cite{Shiralilou:2019div}.

 In Ref.~\cite{Piattella:2017uat}, the two aforementioned observables were combined and the possibility of using them to study cosmology discussed. The authors however, similarly to Sandage and McVittie back in the 60', concluded that the magnitude of the drift effects is so small that it is not detectable with present observations.

In this letter, we suggest a method to test the equivalence principle from constraints on the time variation of the time delay. 
Indeed, as we will discuss later, while spectroscopic measurements of the redshift drift are blind to the variations of $G$, the redshift drift of photons passing through the gravitational potential generated by the lens carry information on the time dependence of the effective gravitational coupling.   

Several theories of modified gravity and Dark Energy imply the violation of the equivalence principle at some level, see for example Ref.~\cite{Bonvin:2018ckp}. For a review of the constraints on the variation of the fundamental constants of nature, see for example Ref.~\cite{Uzan:2010pm}.
Usually, constraints on the variation of $G$ come from experiments performed on galactic and solar system scales, like the Lunar Laser Ranging \cite{Hofmann:2018myc} and pulsars timing \cite{Verbiest:2008gy,Goldman} to mention some of them. Other constraints are obtained by noting that the stellar evolution would be modified by a varying $G$, with an impact for example on the age of  old globular clusters Ref.~\cite{DeglInnocenti:1995hbi}, or from Big Bang Nucleosynthesis \cite{Barrow1978}. The former could be considered as very low redshift constraints, while the latter as high redshift constraints. An interesting constraint at intermediate redshift could be extrapolated by the luminosity of SNIa \cite{Gaztanaga:2001fh} by modifying their Chandrasekhar mass.

The main purpose of this work is to show how time delay measurements provide another way of constraining the variation of $G$ at the redshift of the lens and thus, similarly to SNIa, can be used to test the equivalence principle at intermediate scales. 
Our conclusion is that current observations produce constraints which are several order of magnitude weaker than the constraints existing in the literature, but are likely to improve in the future.

The structure of the paper is the following: in section~\ref{sec1} we briefly review the results of \cite{Piattella:2017uat} and discuss how the time evolution of a point mass lens system is affected by the presence of a time-dependent gravitational coupling. In  section~\ref{sec2} we discuss the application of the above scheme to a realistic model of lens. In section~\ref{sec3} we use both simulated data and light curves of the lensed system DES J0408-5354 to illustrate the method. Finally, section~\ref{concl} is devoted to a discussion of the results and our conclusions.

Through the paper we will indicate the Newton constant with $G_N$ and the general effective gravitational coupling with $G$, assuming that the second can vary in time.

\section{Redshift drift of gravitational lensing for a point mass}\label{sec1}
The basic ingredient of gravitational lensing is the lens equation, which in the thin lens approximation for a point mass reads \cite{Weinberg:2008zzc} :
\begin{equation}\label{LEPM}
   \beta_E^2\equiv \beta\left(\beta - \alpha\right) = \frac{4G_N M}{\mathcal{D}_L }\frac{\mathcal{D}_{LS}}{\mathcal{D}_S} =4G_N M\left(1+z_L\right)\left(\frac{1}{\chi_L} - \frac{1}{\chi_S}\right) \; ,
\end{equation}
where the $\mathcal{D}$'s are the angular diameter distances, $\beta$ is the angular position of the image, $\alpha$ the angular position of the source while $z_L$ indicates the redshift of the lens and $\chi_L$, $ \chi_S$ the comoving distance to the lens and to the source.

The redshift, due to the Hubble flow, is a time-dependent quantity and we can define the redshift drift as:
\begin{equation}\label{RDdef}
    \frac{d z}{d t}= H_0 \left( 1+ z \right) - H(z) \; .
\end{equation}
As discussed in \cite{Piattella:2017uat}, applying Eq.~\eqref{RDdef} to the time derivative of \eqref{LEPM} results in a drift of the angular position of a source and the time delay between incoming signals from the multiple images of a point lens system. The \textit{angular drift} reads:
\begin{equation}\label{ADPM}
    2\frac{\dot{\beta}_E}{\beta_E} = H_0 - \frac{H(z_L)}{1+z_L} \; ,
\end{equation}
while the \textit{time delay drift} reads :
\begin{equation}\label{TDDPM}
\frac{\dot{\Delta}}{\Delta} = K\left(H_0 - \frac{H(z_L)}{1+z_L}\right) \; ,
\end{equation}
where the time delay $\Delta = \Delta_{pot} + \Delta_{geo}$ is defined as the sum of the geometric and the potential time delay, and $K$ is a constant factor, generally of order 1, that depends on the mass of the lens and angles relative to the multiple images, see for example section 9.4 of Ref.~\cite{Weinberg:2008zzc} for more details.
\subsection*{Drift effects in presence of a time-dependent gravitational coupling}
Equations \eqref{ADPM} and \eqref{TDDPM} are obtained assuming that the only time-dependent quantity in Eq.~\eqref{LEPM} is the redshift of the lens $z_L$. However, in several alternative theories of gravitation the gravitational coupling evaluated in the quasi-static approximation (QSA) for scales deep inside the Hubble radius becomes a dynamical quantity and, as a consequence,  the Newton constant $G_N$ must be replaced by a time-dependent effective coupling we denote $G$.  In a generic $f(R)$ theory, for example, we have $G \sim 4G_N/3f_R $, so that $G$ acquires a time dependence from the derivative of $f$ with respect to $R$. For a detailed discussion and extension to more general cases see Ref.~\cite{DeFelice:2011hq}. 
In the deflection angle due to the gravitational lensing effect, we have a degeneracy between the mass of the lens and the gravitational coupling $G$, thus it would not be possible in principle to distinguish a stronger coupling from a heavier mass.
On the other hand, the situation is different if we take into account drift effects. Indeed, in this case Eqs.~\eqref{ADPM} and \eqref{TDDPM} gain an extra contribution:
\begin{equation} \label{ADPMG}
    2\frac{\dot{\beta}_E}{\beta_E} = H_0 - \frac{H(z_L)}{1+z_L}  +\frac{\dot{G}}{G} \; ,
\end{equation}
\begin{equation} \label{TDDPMG}
\frac{\dot{\Delta}}{\Delta} = K\left(H_0 - \frac{H(z_L)}{1+z_L} +\frac{\dot{G}}{G}\right)\; ,
\end{equation}
where it is understood that $G$ is the gravitational coupling at the redshift of the lens.
The above equations show explicitly that drift effects in the context of strong lensing, contrary to what happens with spectroscopic measurements, are sensitive to variations of the gravitational coupling.

As argued in \cite{Piattella:2017uat}, a signal of order $H_0 - H_L/(1+z_L)$, assuming a realistic cosmological evolution, is beyond the sensitivity of current observations. This, in turns, implies that the non-detection of such a signal could be used to put upper bounds on the variation of $G$, thus employed as a test of the equivalence principle. Since time delay measurements are generally more precise than angular ones, in the next section we will discuss how to extend the above framework to the time delay drift of more realistic lensing profiles beyond the point mass approximation.

\section{Time delay drift for an extended lens profile}\label{sec2}
In the thin lens approximation the lens equation for a general mass distribution can be written as:
\begin{equation}\label{LE}
    \left(\bm{\beta}- \bm{\alpha} \right) = \nabla_{\bm{\theta}}\psi\left(\bm{\beta}\right) \; ,
\end{equation}
where $\bm{\beta} = (\beta_1,\beta_2)$ and $\bm{\alpha} = (\alpha_1,\alpha_2)$ are the position in the sky  of the image and the source respectively, and $\nabla_{\theta}$ is the two-dimensional angular gradient.
The quantity $\psi(\bm{\beta})$ appearing in Eq.~\eqref{LE} is the \textit{lensing potential} and is defined as:
\begin{equation}\label{LP}
    \psi(\bm{\beta}) \equiv \frac{2}{c^2 }\frac{\mathcal{D}_{LS}}{\mathcal{D}_L\mathcal{D}_S}\int_{\bm{\beta}}  d\lambda \; \Phi \; ,
\end{equation}
where $\Phi$ is the standard Newtonian gravitational potential and the integral is taken along the path of the light ray, which depends on $\bm{\beta}$ and is  parametrized by $\lambda$.
Taking the divergence of Eq.~\eqref{LE}, as long as the extent of the lens is small compared to cosmological distances, we can use the Poisson equation to relate the Laplacian of the lensing potential to the mass distribution of the lens:
\begin{equation}\label{LapPsi}
\nabla^2_{\bm{\theta}}\psi\left(\bm{\beta}\right) = \frac{8\pi G_N}{c^2}\frac{\mathcal{D}_L\mathcal{D}_{LS}}{\mathcal{D}_S} \Sigma(\bm{\beta}) \; ,  
\end{equation}
where we have defined the surface mass density:
\begin{equation}\label{SMD}
    \Sigma (\bm{\beta}) \equiv \int_{\bm{\beta}} d\lambda  \;\rho \;,
\end{equation}
in which appears the mass distribution of the lens $\rho$. For a detailed derivation and an explanation on the assumptions behind Eqs.~\eqref{LE}, \eqref{LP}, \eqref{LapPsi} see for example Ref.~\cite{Bartelmann:2016dvf}. 

The time delay between two multiple images $i,j$ for a generic lens can be  written, see  Ref.~\cite{Suyu:2012aa}, in the form:
\begin{equation}\label{TD}
    \Delta_{ij} = \frac{D_{\Delta_t}}{c}\left( \frac{\left(\bm{\beta}_i - \bm{\alpha}\right)^2}{2} - \frac{\left(\bm{\beta_j} - \bm{\alpha}\right)^2}{2} + \psi\left(\bm{\beta_j}\right) - \psi\left(\bm{\beta_i}\right)\right)\; ,
\end{equation}
where it was defined the time delay distance:
\begin{equation}\label{timedelaydistance}
    D_{\Delta_t} \equiv \left(1+ z_L\right)\frac{\mathcal{D}_L \mathcal{D}_S}{\mathcal{D}_{LS}} \; .
\end{equation}
Inside the brackets, one can distinguish the geometric time delay contribution in the first two terms, which accounts for the delay induced by the different lengths of the light rays trajectories, and the potential one in the last two, which instead accounts for the delay induced on photons while they travel through the gravitational potential of the lens. The above expression of the time delay is the one usually employed by the CosmoGrail collaboration to quantify the shifting of the light curves of different images from strongly lensed systems.

We want to evaluate the time derivative of Eq.~\eqref{TD} with the hypothesis of a time-dependent gravitational coupling $G$. Let us begin with the time derivative of the lensing potential Eq.~\eqref{LP}: 

\begin{equation}
    \dot{\psi}(\bm{\beta})= \frac{2}{c^2 }\frac{\mathcal{D}_{LS}}{\mathcal{D}_L\mathcal{D}_S}\left( \frac{1}{\left(1+z_L\right)}\frac{d z_L}{dt}\int_{\bm{\beta}} d\lambda \; \Phi + \frac{d}{dt}\left(\int_{\bm{\beta}} d\lambda \; \Phi \right)\right) \; .
\end{equation}
The time dependence of the second term in the right hand side of the latter equation comes from the change in time of the light ray path due to a variation of $\bm{\beta}$ and to the dependence of the Newtonian potential. The former is difficult to compute exactly because it is in general difficult to describe precisely the curve identified by the light ray path. On the other hand, it is reasonable to assume that the induced variation of the curve is small and does not contribute to the support of the integral in Eq.~\eqref{LP}. In particular, this is true if we evaluate the above integral within the the Born approximation, \textit{i.e.} along the unperturbed light path, as it is customary in strong lensing applications where $\Phi/c^2 \ll 1$  \cite{Bartelmann:2016dvf}. Within this assumption we can interchange the operation of integration and time differentiation obtaining:
\begin{equation}
     \dot{\psi}(\bm{\beta}) = \psi(\bm{\beta})\left(H_0 - \frac{H(z_L)}{1+z_L}\right) + \psi(\bm{\beta})\frac{\int_{\bm{\beta}} d\lambda \; \dot{\Phi}}{\int_{\bm{\beta}} d\lambda\; \Phi} \; .
\end{equation}
The above equation can be further simplified if we consider a static distribution of matter. Indeed, in this case the only time dependence of the Newtonian potential is through the effective gravitational coupling $G$,  so that we have:
\begin{equation}\label{TDTDSD}
\dot{\psi}(\bm{\beta}) = \psi(\bm{\beta})\left(\frac{\dot{G}}{G} + H_0 - \frac{H(z_L)}{1+z_L}\right) \; .
\end{equation}
In concrete time delay measurements, the time delay distance  must be corrected in order to take into account the effect of the mass distributed along the line of sight and the effect of the peculiar velocity of the lens. This correction is implemented by introducing the external convergence $\kappa_{ext}$, which relates the real time delay distance $D_{\Delta t}^{real}$ to $D_{\Delta t}$ via:
\begin{equation}
    D_{\Delta t}^{real} \equiv\frac{D_{\Delta t}}{1 - \kappa_{ext}} \; ,
\end{equation}
see for example Ref.~\cite{Suyu:2012aa}.  If we assume that the external convergence has a time dependence this would be inherited by the time delay distance \eqref{timedelaydistance}.

Finally, by mean of Eq.~\eqref{TDTDSD}, we are able to define the  logarithmic time derivative, i.e. the relative variation, of the time delay:
\begin{equation}\label{TDDk}
    \frac{\dot{\Delta}_{ij}}{\Delta_{ij}} = \left(\frac{\dot{G}}{G}  + H_0 - \frac{H(z_L)}{1+z_L}  \right)\left[1  + \frac{D_{\Delta t}\left(\bm{\beta}_i - \bm{\alpha}\right)^2}{2c \Delta_{ij}} - \frac{D_{\Delta t}\left(\bm{\beta_j} - \bm{\alpha}\right)^2}{2c \Delta_{ij}}\right] -\frac{\dot{\kappa}_{ext}}{1-\kappa_{ext}} \; .
\end{equation}

It is interesting to note that the term due to the external convergence does not depend on the angular position of the images, contrary to the one inside square brackets. Thus, even if not constant, it is the same for all the multiple images in a lensed system, and considering differences of their time delay drift will allow to eliminate its contribution.

\section{Estimating the variation of the time delay from current data}\label{sec3}

We now illustrate how the data can put constraints on the variation of the time delay. To this end, we use the package PyCS3 from the COSMOGRAIL program \footnote{Available  \href{http://cosmograil.org/}{here}.}, see Refs.~\cite{Tewes:2012gs,Bonvin:2015jia}. The data used are the simulated light curves used in \cite{Bonvin:2015jia}, produced in the context of the blind time delay measurement
competition named Time Delay Challenge 1 (TDC1) \cite{Dobler:2013rda}, and from the quasar DES J0408-5354 \cite{Courbin:2017yvz}. We split the total time of observations in two equal time periods. Each period consists of 658 days for the trial curves, and of 93 days for DES J0408-5354. The time delay between each image is then calculated for each period. The time delay estimates are shown in App.~\ref{appendix} and summarized in Table \ref{tabletimedelay}. From it we can readily estimate the relative variation \eqref{TDDk} as:
\begin{equation}
     \frac{\dot{\Delta}_{ij}}{\Delta_{ij}}= \frac{\Delta_{ij}\left(t + \delta t\right) - \Delta_{ij}(t)}{\delta t \Delta_{ij}(t)} = \frac{\Delta_{ij}^{I+II} - \Delta_{ij}^{I}}{\Delta_{ij}^{I} \delta t} \; ,
\end{equation}
We display the results in Table \ref{tableredshiftdrift}. 

\begin{table}
    \centering
    \begin{tabular}{c||c|c|c|c|c|c}
          & $\Delta_{AB}$ & $\Delta_{AC}$ & $\Delta_{AD}$ & $\Delta_{BC}$ & $\Delta_{BD}$ & $\Delta_{CD}$ \\
         \hline
         Trial I & $-4.9^{+4.3}_{-5.5}$ & $-18.0^{+6.0}_{-8.0}$ & $-0.7^{+9.6}_{-10.3}$ & $-13.5^{+7.6}_{-7.6}$  & $+4.3^{+9.8}_{-11.5}$ & $+17.5^{+10.7}_{-12.1}$  \\
         Trial I+II & $-4.5^{+2.4}_{-2.0}$ & $-21.3^{+1.4}_{-1.7}$ & $-29.9^{+9.2}_{-7.0}$ & $-16.6^{+1.8}_{-3.4}$  & $-25.2^{+7.5}_{-6.4}$ & $-8.2^{+8.3}_{-6.7}$ \\
         \hline
         DES 0408 WFI I & $-106.5^{+15.9}_{-14.1}$ & $-110.6^{+50.4}_{-21.7}$ & $-142.2^{+34.3}_{-18.3}$ & $-2.4^{+38.8}_{-19.9}$  & $-37.1^{+25.7}_{-16.2}$ & $-31.7^{+19.8}_{-30.3}$ \\
         DES 0408 WFI I+II & $-112.6^{+6.6}_{-2.2}$ & $-117.2^{+5.9}_{-7.6}$ & $-153.2^{+11.9}_{-9.5}$ & $-7.1^{+8.6}_{-8.6}$  & $-40.5^{+11.3}_{-9.3}$ & $-35.6^{+14.1}_{-10.7}$
    \end{tabular} 
    \caption{Time delay $\Delta$ between four images from a simulated quasar and the DES J0408-5354 quasar \cite{Courbin:2017yvz}. Each image is labeled from $A$ to $D$.  The numbers I and I+II indicate that $\Delta$ was measured over the first half of the period of observations, or over the whole period, respectively. All values are given in days.}
    \label{tabletimedelay}
\end{table}

\begin{table}
    \centering
    \begin{tabular}{c||c|c|c|c|c|c}
          &$|\dot{\Delta}/\Delta|_{AB}$ & $|\dot{\Delta}/\Delta|_{AC}$  & $|\dot{\Delta}/\Delta|_{AD}$& $|\dot{\Delta}/\Delta|_{BC}$& $|\dot{\Delta}/\Delta|_{BD}$ & $|\dot{\Delta}/\Delta|_{CD}$\\
         \hline
         Trial ($\times 10^{-5}$) & X & $27.9 \pm 12.6$ & X & $34.9 \pm 20.9 $ & X & X \\
         \hline
         DES 0408 WFI ($\times 10^{-5}$) & $61.6 \pm 9.9$ & $64.2 \pm 29.6$ & $83.2 \pm 21.1$ & X & $98.5 \pm 73.6$ & X
    \end{tabular}  
    \caption{Estimated absolute time delay variation for the simulated quasar and DES J0408-5354. All values are given in $day^{-1}$. The X's represent values with uncertainty bigger than the central value, and are thus omitted.}
    \label{tableredshiftdrift}
\end{table}

\subsection*{Estimated constraint on the gravitational constant variation}

Through Eq.~\eqref{TDDk} it is possible to relate constraints on the relative time variation of $\Delta_{ij}$ to upper bounds on the variation of $\dot{G}/G$. We can indeed estimate the magnitude of the redshift drift effect due to the Hubble flow to be of the order of $H_0 \sim 10^{-10} \, yr^{-1}$, so we assume that its effect is negligible within the  precision of the measurements. 
The external convergence time dependence is difficult to evaluate. We expect it to depend explicitly on $\dot{G}/G$, similarly to the potential generated by the lens, with the two contributions having the same sign, but in what follows we will assume that it is negligible with respect to the precision of current observations. We want to stress however,  as we discussed in the previous sections, that it is in principle possible to disentangle such contribution by considering differences of time delays drifts of multiple images.

Another effect that could be of relevance is the time variation of the time delay due to the peculiar velocity of the lens galaxy and its transverse motion. On the other hand, according to Refs.~\cite{Zitrin:2018let,Wucknitz:2020spz}, this contribution is estimated to be of the order of a few seconds per year, so that the effect, even though it is bigger than the cosmological one due to the Hubble flow, is still not appreciable with current precision. Under these assumptions, and roughly estimating  the term inside the square bracket of Eq.~\eqref{TDDk} to be of order 1,  the values of Table \ref{tableredshiftdrift} can be directly converted into upper bounds on the time variation of the effective gravitational coupling. The results are reported in Table \ref{tableGconst}.
\begin{table}
    \centering
    \begin{tabular}{c||c|c|c|c|c|c}
          &$|\dot{G}/G|_{AB}$&
          $|\dot{G}/G|_{AC} $&
          $|\dot{G}/G|_{AD} $& 
          $|\dot{G}/G|_{BC} $&
          $|\dot{G}/G|_{BD} $&
          $|\dot{G}/G|_{CD} $\\
         \hline
         Trial ($\times 10^{-1}$) & X & $1.0  \pm 0.5$ & X & $1.3 \pm 0.8 $& X & X \\
         DES 0408 WFI ($\times 10^{-1}$) & $2.2 \pm 0.4$& $2.3 \pm 1$ &  $3.0 \pm 0.8$ & X & $3.6\pm 3$ & X
    \end{tabular}  
    \caption{Upper bounds on the absolute value of $\dot{G}/G$ in $yr^{-1}$ from the simulated quasar and DES J0408-5354.} 
    \label{tableGconst}
\end{table}
\section{Summary and Discussion}\label{concl}
The main purpose of this work is to illustrate how time delay measurements can be employed as a test of the equivalence principle. More specifically, we show that the variation of the time delay is an observable sensitive to the time variation of the gravitational coupling $G$. In the case of a lens with a static mass distribution, we were able to write explicitly such a dependence in Sec.~\ref{sec2}, \textit{i.e.} Eq.~\eqref{TDDk}. We also showed that in principle it is possible, in a system with multiple lensed images, to disentangle the contribution of the external convergence time variation by considering differences in the time delay variation of multiple images. In Sec.~\ref{sec3}, for illustrative purposes, we considered both a simulated dataset and the real light curves of the lens system DES J0408-5354 to study how the data could constrain the relative variation of the time delay. We split the total observation time in two data sets and considered how the estimated time delay of each image changed in the two epochs. 
Then, assuming that the contributions from time delay drift, transverse motion and variation of the external convergence are, as expected by theoretical predictions, beyond the sensitivity of current experiments, we converted the estimated variation of the time delay in upper bounds on the variation of the gravitational coupling $|\dot{G}/G|$.
The constraint, which in the most optimistic cases are of order $\sim 10^{-1}-10^{-2}\, yr^{-1}$,  are $\sim 10$ order of magnitude weaker than the ones obtained with other probes, which are in general of order $10^{-11}-$ $10^{-12}$. At the moment, these bounds are clearly too weak to be competitive with the bounds from other probes, or to be employed to constrain modified gravity theories. On the other hand, if one day they will be realizable with sufficient precision, they would have the advantage of probing extragalactic scales, which are usually unexplored in this context. We want to stress that the constraints derived here can improve in several ways, in particular with a prolonged period of observation. The light curves of the quasar DES J0408-5354, for example, were obtained over a period of roughly 7 months. A similar variation of the time delay, of the order of days in the best scenario, over a period of 4 years would improve the constraint of a factor 10. We also expect more precise data, like the light curves of Ref.~\cite{Millon:2020xab}, to result also in a constraint at least one order of magnitude stronger. Finally, as discussed in Refs.~\cite{Zitrin:2018let,Wucknitz:2020spz,Liu:2019jka}, if in the future  strongly lensed repeating FRB are detected, they will provide time delay measurements of such extremely high precision, nominally of the order of seconds, that even redshift drift effects due to the Hubble expansion will be appreciable. In this scenario, one would need to study how to distinguish between a single repeated FRB and its lensed images, but a treatment of this issue goes beyond the scope of this work.

Strong lensing time delay measurements are an extremely valuable tool to study our Universe which already provide a competitive model-independent probe to investigate post-Newtonian and cosmological parameters. The goal of this paper is to suggest a novel instrument for testing the equivalence principle on extragalactic scales. At the moment, the contribution of  time delays variations in this regard is very modest, but we keep high expectations for the future.

\acknowledgments

We are grateful to Oliver Piattella, Matteo Martinelli and Rodrigo Von Marttens for valuable comments and discussions. We are also extremely grateful to Martin Millon for discussions about PyCS3. This study was financed in part by the Coordenação de Aperfeiçoamento de Pessoal de Nível Superior - Brasil (CAPES) - Finance Code 001.

\appendix

\section{Estimating Uncertainties with PyCS3}
\label{appendix}

\begin{figure}[h!]
    \centering
    \includegraphics[scale=0.4]{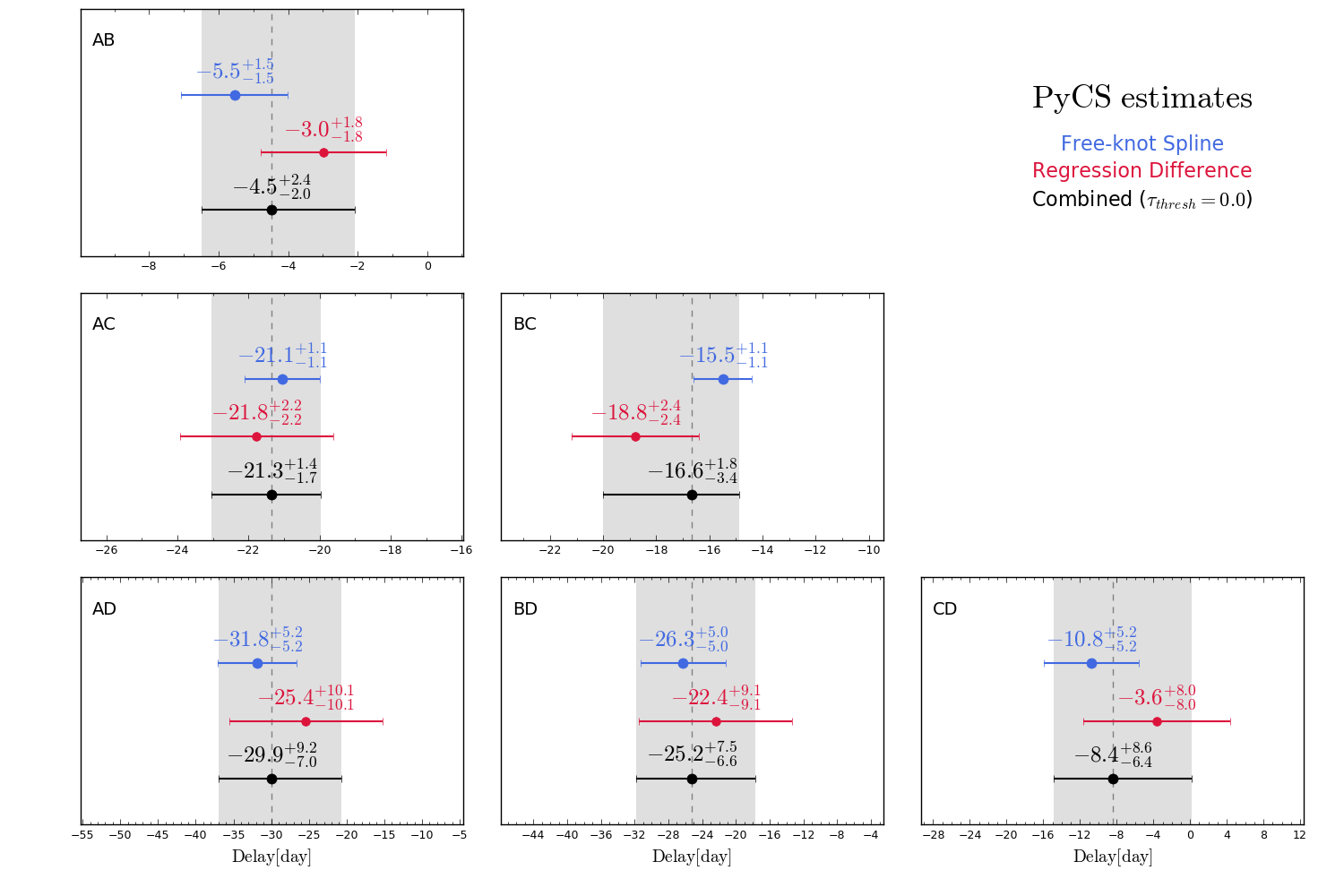}
    \includegraphics[scale=0.45]{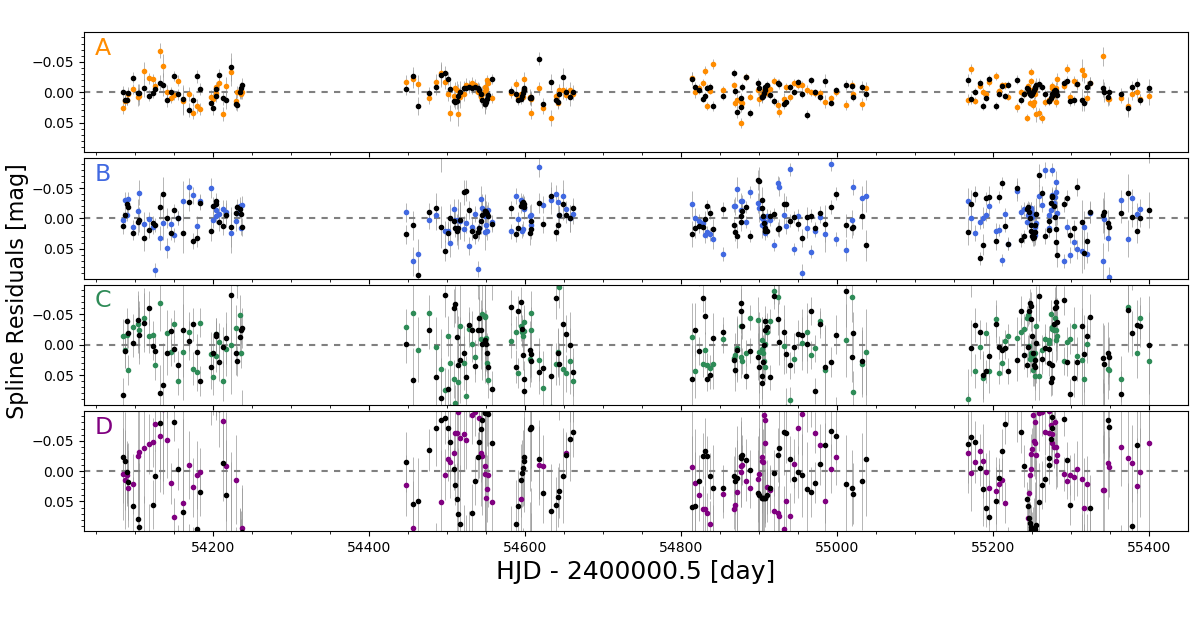}
    \includegraphics[scale=0.45]{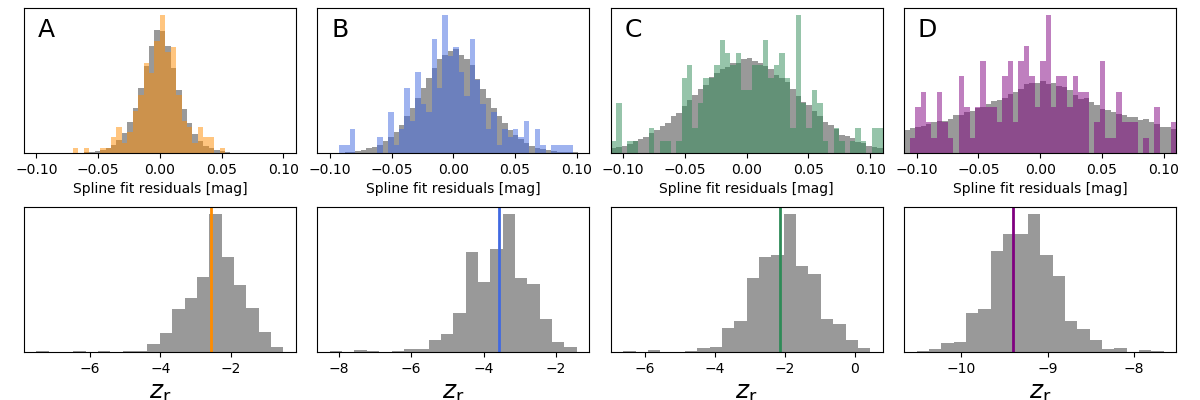}
    \caption{Time delay estimates for the simulated quasar over the full observation period.}
    \label{fig:simul}
\end{figure}

\begin{figure}[htp!]
    \centering
    \includegraphics[scale=0.4]{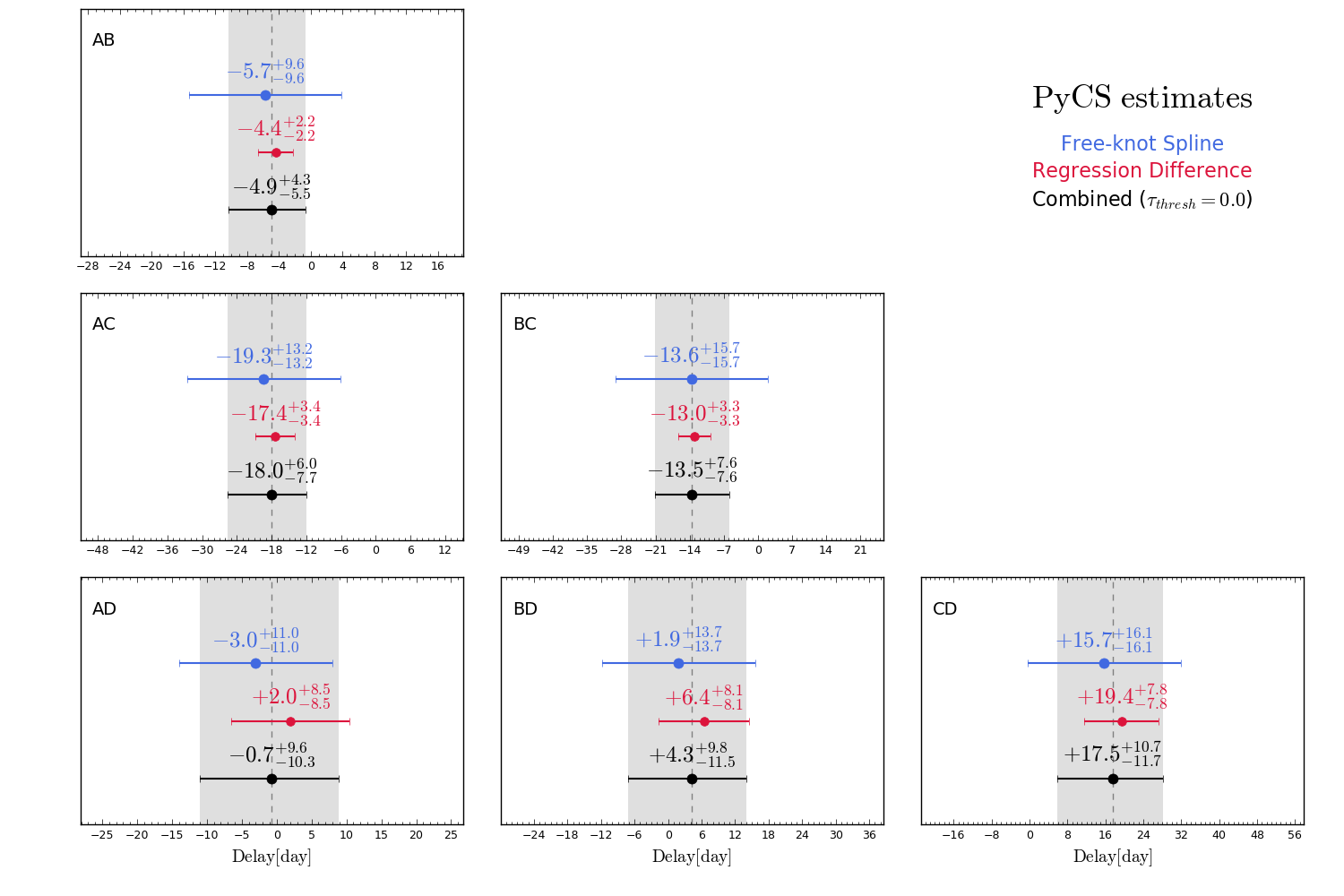}
    \includegraphics[scale=0.4]{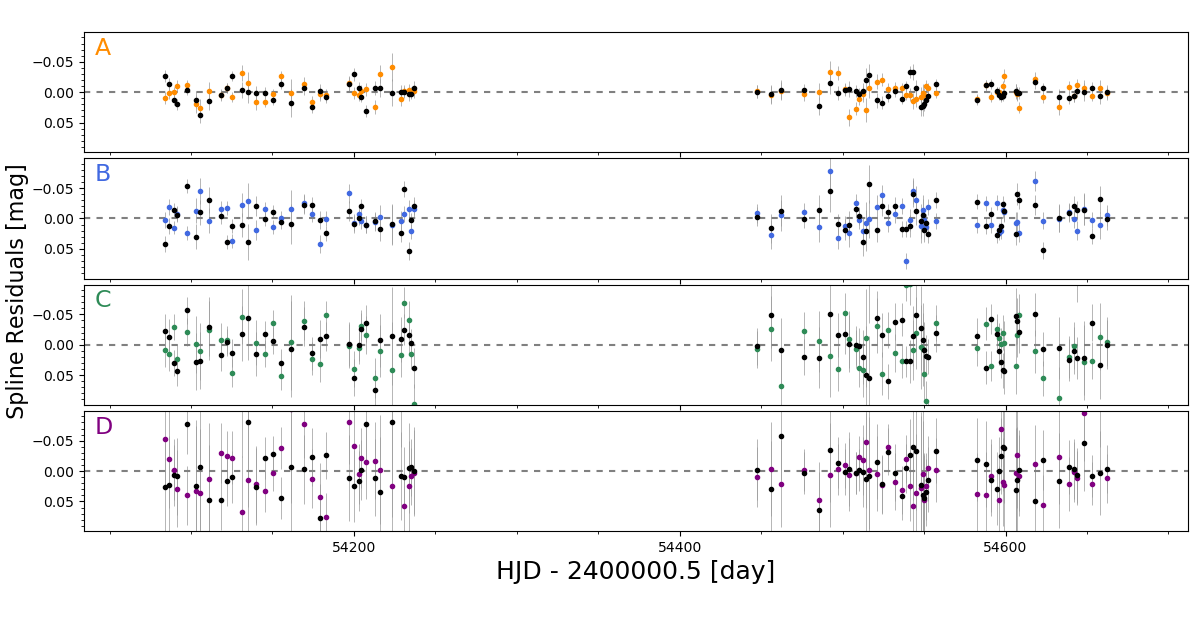}
    \includegraphics[scale=0.4]{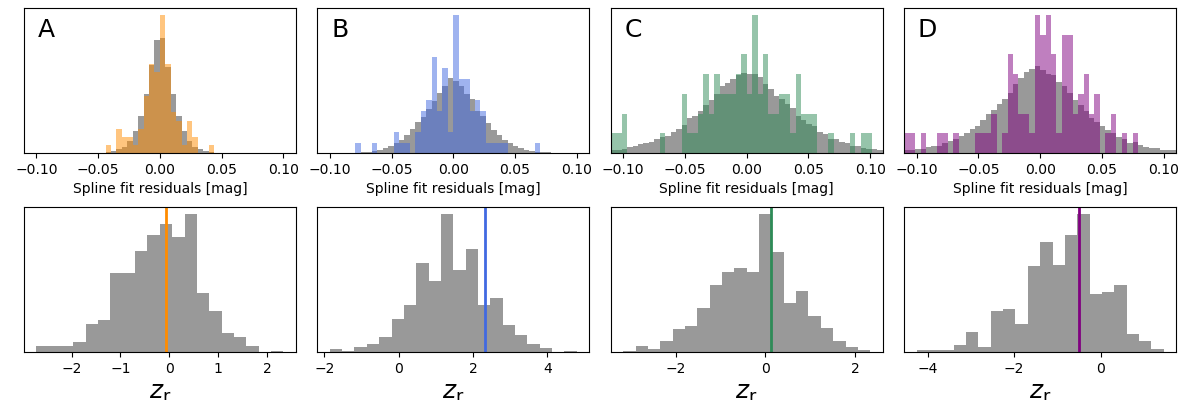}
    \caption{Time delay estimates for the simulated quasar over half of the observation period.}
    \label{fig:simulhalf}
\end{figure}

\begin{figure}[htp!]
    \centering
    \includegraphics[scale=0.4]{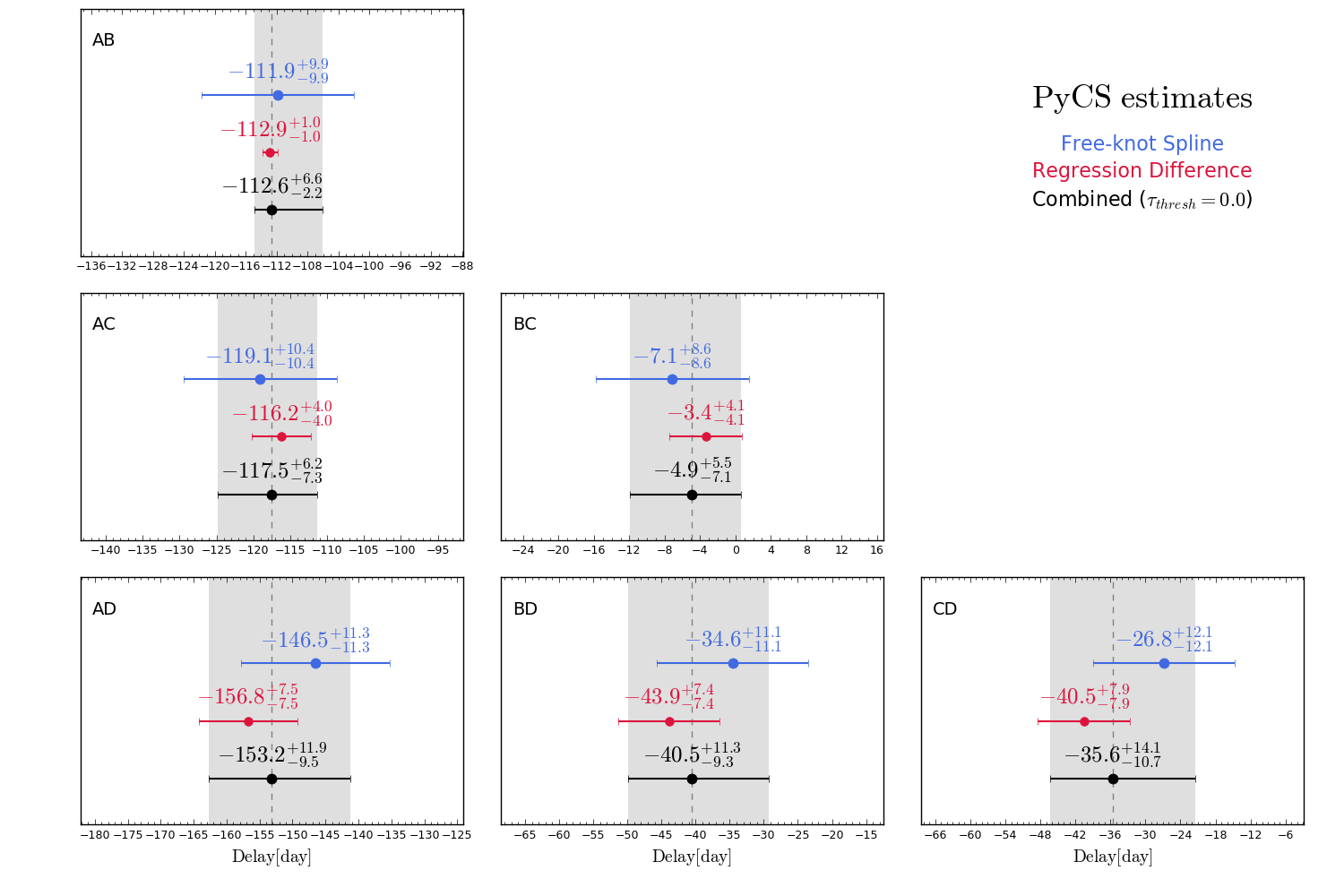}
    \includegraphics[scale=0.4]{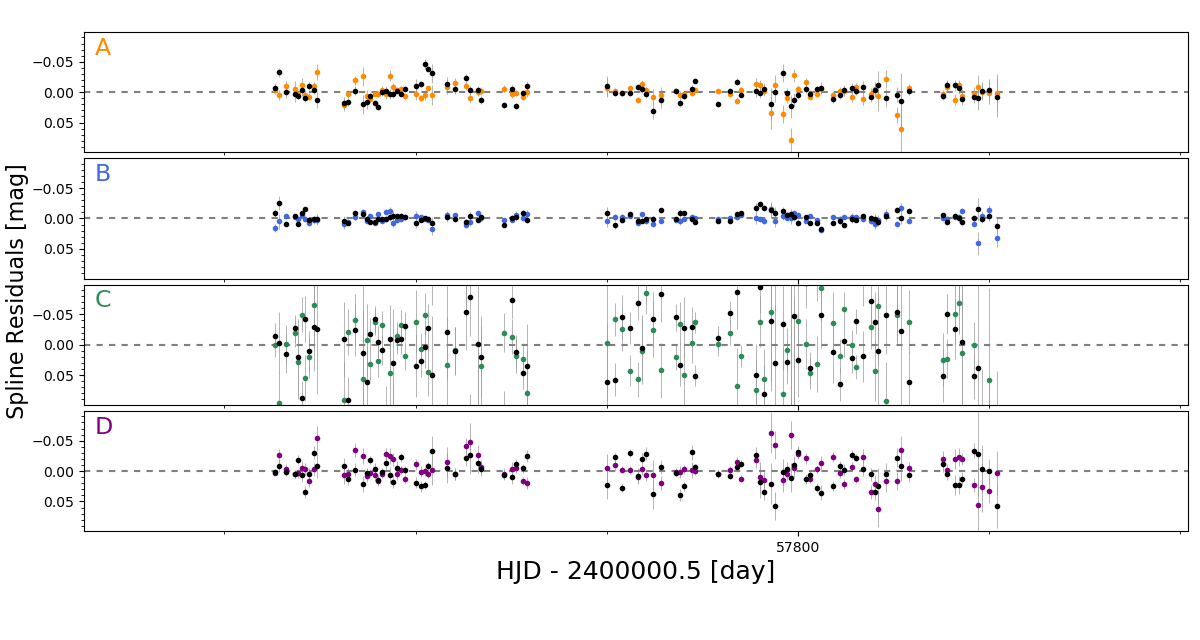}
    \includegraphics[scale=0.4]{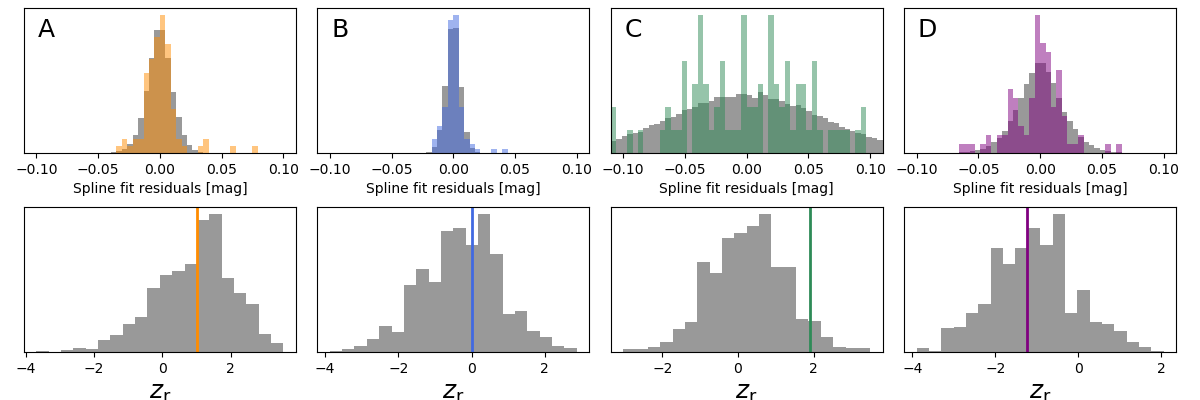}
    \caption{Time delay estimates for the quasar DES J04078-5354 over the full observation period.}
    \label{fig:des}
\end{figure}

\begin{figure}[htp!]
    \centering
    \includegraphics[scale=0.4]{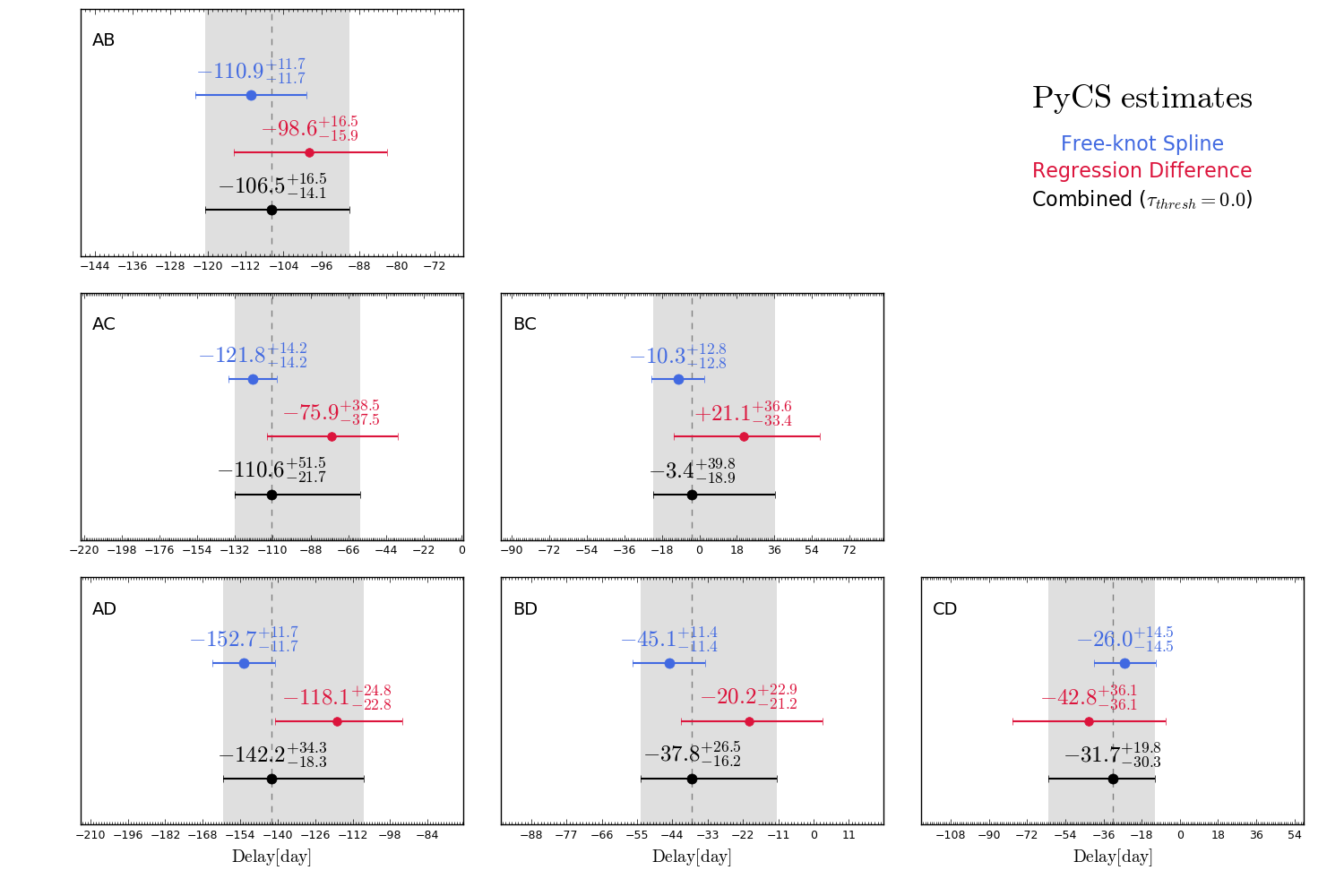}
    \includegraphics[scale=0.4]{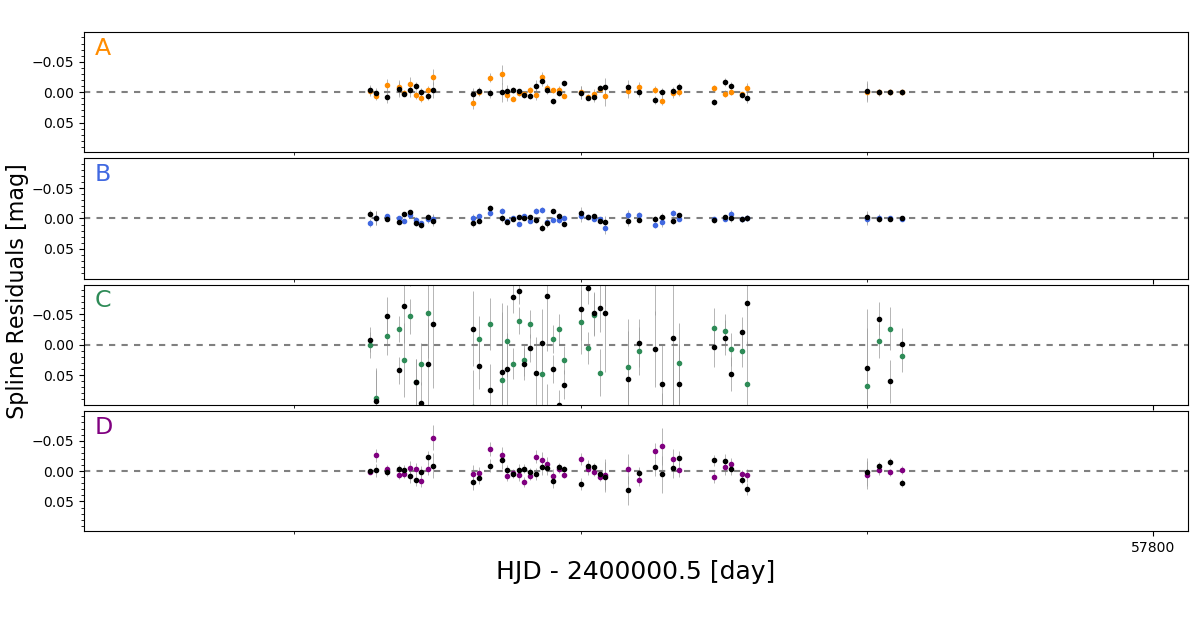}
    \includegraphics[scale=0.4]{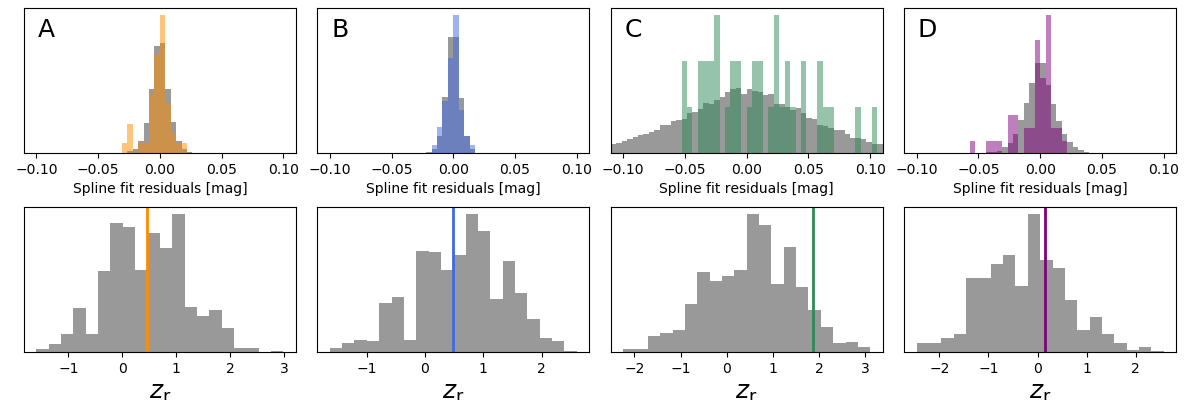}
    \caption{Time delay estimates for the quasar DES J04078-5354 over half of the observation period.}
    \label{fig:deshalf}
\end{figure}

In this appendix, we display the time delays and their uncertainties obtained with PyCS3. Uncertainties are obtained by simulating light curves close to the data, and randomizing the true time delay applied to each curve, see Sec.~(3.2) of Ref.~\cite{Millon:2020xab} for more details. The final marginalization is done performing a hybrid approach between the ``free-knot spline" and the ``regression difference" estimators as explained in Sec.~(3.3) of Ref.~\cite{Millon:2020xab}. The parameter $\tau_{thresh}=0$ indicates the marginalization is done over the two estimators. In each figure, the top panels show the final time delay estimates marginalizing over the two estimators. The middle figure shows the residuals for the spline fit to the data. The top row of the bottom panels show the distribution of data residuals for mock curves (in gray) and data (in colors), whereas the bottom panels show their normalization over the number of runs $z_r$. The time delay estimates for the simulated curves over the whole period of observations ($\approx 1.316$ days) is shown in Fig.~\ref{fig:simul} while Fig.~\ref{fig:simulhalf} is over half the total period. Similarly, Fig.~\ref{fig:des} ($\approx 189$ days) and Fig.~\ref{fig:deshalf} show time delays for the object DES J0408-5354.

\clearpage
\bibliographystyle{unsrturl}
\bibliography{TDD.bib}
\end{document}